\documentclass[amsmath,useAMS,usenatbib]{mn2e}
\usepackage{epsfig}
\usepackage{graphicx}
\usepackage{amsmath}
\usepackage{color}
\usepackage{psfrag}
\def\d{{\rm d}}

\newcommand{\pderiv}[2]{\frac{\partial #1}{\partial #2}}
\newcommand{\deriv}[2]{\frac{d #1}{d #2}}
\newcommand{\laplace}{{ \nabla}^2}
\newcommand{\ddt}[1]{\frac{\partial ^2 #1}{\partial t^2}}
\newcommand{\ddS}[1]{\frac{d ^2 #1}{d S^2}}
\newcommand{\Jp}{{\bf J}_p}
\newcommand{\del}{\nabla}

\newcommand{\curl}[1]{{ \nabla}\times #1}
\newcommand{\dds}[1]{\frac{{\rm d} #1}{{\rm d}S}}
\def\rmmat#1{{\hbox{\rm #1}}}

\def\d{{\rm d}}
\newcommand{\bfi}{{\bf B}}
\newcommand{\efi}{{\bf E}}

\newcommand{\bhat}[1]{\hat{\bf #1}}
\begin{document}
\title{Nonlinear Electromagnetic Waves in Magnetosphere of a Magnetar}
\date{Accepted ---.  Received ---; in original form ---}
\author[D. Mazur and J. S. Heyl]{Dan Mazur and Jeremy
  S. Heyl$^{\dagger}$ \\
Department of Physics and Astronomy, University of British Columbia, 
Vancouver, British Columbia, Canada, V6T
  1Z1 \\
$^{\dagger}$Email: heyl@phas.ubc.ca; Canada Research Chair
}
\pagerange{\pageref{firstpage}--\pageref{lastpage}} \pubyear{2010}

\maketitle
\label{firstpage}

\begin{abstract}
  We compute electromagnetic wave propagation through the
  magnetosphere of a magnetar.  The magnetosphere is modeled as the
  QED vacuum and a cold, strongly magnetized plasma.  The background
  field and electromagnetic waves are treated nonperturbatively and
  can be arbitrarily strong.  This technique is particularly useful
  for examining non-linear effects in propagating waves. Waves
  travelling through such a medium typically form shocks; on the other
  hand we focus on the possible existence of waves that travel without
  evolving.  Therefore, in order to examine the nonlinear effects, we
  make a travelling wave ansatz and numerically explore the resulting
  wave equations.  We discover a class of solutions in a 
  homogeneous plasma which are
  stabilized against forming shocks by exciting nonorthogonal
  components which exhibit strong nonlinear behaviour. These waves may
  be an important part of the energy transmission processes near
  pulsars and magnetars.
\end{abstract}
 \maketitle
\section{Introduction}
\label{sec:introduction}
The magnetosphere of a magnetar provides a particularly interesting
medium for the propagation of electromagnetic waves.  Magnetars are
characterized by exceptionally large magnetic fields that can be
several times larger than the quantum critical field
strength~\citep{mereghetti-2008}.  Because the magnetic fields are so
large, the fluctuations of the vacuum of quantum
electrodynamics (QED) influence the propagation of light.
Specifically, the vacuum effects add nonlinear terms to the wave
equations of light in the presence large magnetic fields.  In
addition, the magnetosphere of a magnetar contains a plasma which alters
the dispersion relationship for light.  Because of the unique optical
conditions in the magnetospheres of magnetars, they provide excellent
arenas to explore nonlinear vacuum effects arising due to quantum
electrodynamics.

The influence of QED vacuum effects from strong magnetic fields in the
vicinities of magnetized stars has previously been studied by several
authors.  The combined QED vacuum and plasma medium is discussed in
detail in the context of neutron stars in \citet{meszarosbook}.  Vacuum
effects have been found to dominate the polarization properties and
transport of X-rays in the strong magnetic fields near neutron stars
\citep{PhysRevD.19.3565, PhysRevLett.41.1544, 1981ApJ...251..695M,
  1980ApJ...238.1066M, 1983ZhETF..84.1217G}.  Detailed consideration
of magnetic vacuum effects is therefore critical to an understanding
of emissions from highly magnetized stars.

Most studies of waves in systems including plasmas or vacuum effects
approach the problem perturbatively, which limits the applicability of
their results.  The purpose of the present paper is to examine the
combined impact of the QED vacuum and a magnetized plasma using
nonperturbative methods to fully preserve the nonlinear interaction
between the fields.  Neutron stars may be capable of producing very
intense electromagnetic waves, comparable to the ambient magnetic
field.  For example, a coupling between plasma waves and seismic
activity in the crust could produce an Alfv\'en wave with a very large
amplitude~\citep{1989ApJ...343..839B,1995MNRAS.275..255T}. Even
  if they may not be produced directly, electromagnetic waves
  naturally develop through the interactions between Alfv\'en waves.
  To lowest order in the size of the wave, Alfv\'en waves do not
  suffer from shock formation \citep{Thom98} whereas electromagnetic
  waves do \citep{heyl1998electromagnetic,Heyl98mhd}; therefore, how
  to stabilise the propagation of the the latter is the focus of this
  paper.

If such a magnetospheric disturbance results in electromagnetic waves
of sufficiently large amplitude and low frequency, then
nonperturbative techniques are required to characterize the wave.  The
importance of studying such a system nonperturbatively is particularly
well illustrated by the fact that some nonlinear wave behaviour is
fundamentally nonperturbative, as is generally the case with
solitons~\citep{rajaraman1982solitons}.  In order to handle the
problem nonperturbatively, we choose to study waves whose spacetime
dependence is described by the parameter $S=x-vt$, where $v$ is a
constant speed of propagation through the medium in the
$\bhat{x}$-direction.  In the study of waves, one normally chooses the
ansatz $e^{i(\omega t - {\bf k}\cdot{\bf x})}$.  However, in this
picture, a numerical study would typically treat the self-interactions
of the electromagnetic field by summing the interactions of finitely
many Fourier modes.  So, this ansatz conflicts with our goal of
studying the nonlinear interactions to all orders.  In contrast, a
plane wave ansatz given by $S=x-vt$ allows us to study a simple wave
structure to all orders without any reference to individual Fourier
modes.


We model a magnetar atmosphere by including the effects of arbitrarily
strong electromagnetic fields using a QED one-loop effective
Lagrangian approach.  These effects are discussed in
section~\ref{sec:tensors}.  Plasma effects are included by assuming
free electrons moving under the Lorentz force without any
self-interactions.  The model is that of a cold magnetohydrodynamic
plasma and is discussed in section~\ref{sec:plasma}.  We have
  also assumed that the medium is homogeneous in agreement with the
  travelling-wave ansatz.  Of course the actual situation is more
  complicated with a thermally excited plasma \citep[e.g.][]{Gill09BW}
  and inhomogeneities --- the latter can result in a whole slew of
  interesting interactions between the wave modes
  \citep{Heyl99polar,Heyl01qed,Heyl01polar,2003PhRvL..91g1101L} that
  are especially crucial to our understanding of the thermal radiation
  from their surfaces, but these are beyond the scope of this paper.


The formation of electromagnetic shocks is expected to be an important
phenomenon for electromagnetic waves in the magnetized vacuum since
electromagnetic waves can evolve discontinuities under the influence
of nonlinear interactions~\citep{PhysRev.113.1649,
  zheleznyakov1982shock, heyl1998electromagnetic}.  Such shocks can
form even in the presence of a plasma~\citep{PhysRevD.59.045005}.  In
this study, through our explicit focus on travelling waves, 
we examine an alternate class of solutions to the wave equations 
 which do not suffer this fate.  Instead,
they are stabilized against the formation of discontinuities by
nonlinear features.  These waves travel as periodic wave trains
without any change to their form, such as wave steepening or shock
formation.  Waves such as these may contribute to the formation of
pulsar microstructures~\citep{1987STIN...8816622C,2001ApJ...558..302J}.

\section{Wave Equations}
\label{sec:wave-equations}
\subsection{The Maxwell's equations}
\label{sec:maxwells-equations}
The vacuum of QED in the presence of large magnetic fields can be
described as a non-linear optical medium~\citep{1997JPhA...30.6485H}.
We also choose to treat the effect of the plasma on the waves through
source terms $\rho_p$ and $\Jp$; therefore, we begin by considering
Maxwell's equations in the presence of a medium and plasma sources.
In Heaviside-Lorentz units with $c=1$, Maxwell's equations can be used
to derive the wave equations
\begin{eqnarray}
\label{eqn:Maxwell1}
\laplace {\bf D} - \ddt{{\bf D}} &=& -\curl{(\curl{({\bf D}-{\bf E})})}+  \nonumber \\
	& &\pderiv{}{t} (\curl{({\bf B}-{\bf H})}) 
	+{\bf \nabla}\rho_p + \pderiv{{\bf J_p}}{t} \\
\label{eqn:Maxwell2}
\laplace {\bf H} - \ddt{{\bf B}} &=& -\del(\del \cdot({\bf B}-{\bf H}))- \nonumber \\
	& &\pderiv{}{t}(\curl{({\bf D}-{\bf E})}) - \curl{{\bf J_p}}.
      \end{eqnarray}
For clarity, we will avoid making cancellations or dropping vanishing terms.
We define the vacuum dielectric and inverse magnetic permeability tensors as
follows \citep{Jack75}
\begin{equation}
D_i = \varepsilon_{ij} E_j,~~H_i = \mu^{-1}_{ij} B_j.
\end{equation}
In the next few sections we build a model describing travelling waves 
in a magnetar's atmosphere from these equations.

\subsection{Vacuum Dielectric and Inverse Magnetic Permeability Tensors}
\label{sec:tensors}
In this section, we describe our model of the QED vacuum in strong
background fields in terms of vacuum dielectric and inverse magnetic
permeability tensors.  These are most conveniently described in terms
of two Lorentz invariant combinations of the fields.
\begin{eqnarray}
K&=&\left(\frac{1}{2}\varepsilon^{\lambda \rho \mu \nu} 
F_{\lambda \rho}F_{\mu \nu}\right)^2=-(4\efi \cdot \bfi)^2\\
I&=&F_{\mu \nu}F^{\mu \nu} = 2(|\bfi|^2-|\efi|^2)
\end{eqnarray}
In order to examine the nonlinear effects of the vacuum
nonperturbatively, we wish to use vacuum dielectric and inverse
magnetic permeability tensors which are valid to all orders in the
fields.  Analytic expressions for these tensors were derived by \citet{1997JPhA...30.6485H} for the case of wrenchless
fields ($K=-(4\efi \cdot \bfi)^2=0$) from the
Heisenberg-Euler-Weisskopf-Schwinger~\citep{heisenberg-1936-98,
  weisskopf1936kongelige, Schwinger:1951} one-loop effective Lagrangian in
\citet{1997PhRvD..55.2449H} and expressed in terms of a set of analytic
functions.
\begin{eqnarray}
X_0(x) & = & 4 \int_0^{x/2-1} \ln(\Gamma(v+1)) \d v 
+ \frac{1}{3} \ln \left ( \frac{1}{x} \right )\nonumber \\
& &~~+ 2 \ln 4\pi - 4 \ln A-\frac{5}{3} \ln 2 \nonumber \\
& & ~~ - \left [ \ln 4\pi + 1 +  \ln \left ( \frac{1}{x} \right ) \right ] x \nonumber \\
& &~~+ \left [ \frac{3}{4} + \frac{1}{2} \ln \left ( \frac{2}{x} \right )
\right ]
x^2
\label{eqn:x0anal} \\
X_1(x) & = & - 2 X_0(x) + x X_0^{(1)}(x) + \frac{2}{3} X_0^{(2)} (x) -
\frac{2}{9} \frac{1}{x^2}
\label{eqn:x1anal} \\
X_2(x) & = & -24 X_0(x) + 9 x X_0^{(1)}(x)\nonumber \\
& &~~ + (8 + 3 x^2) X_0^{(2)}(x)
+ 4 x X_0^{(3)}(x) \nonumber \\
& & ~~ - \frac{8}{15} X_0^{(4)}(x) + \frac{8}{15}
\frac{1}{x^2} + \frac{16}{15} \frac{1}{x^4} 
\label{eqn:x2anal}
\end{eqnarray}
where
\begin{equation}
X_0^{(n)}(x) = \frac{\d^n X_0(x)}{\d x^n}
\end{equation}
and
\begin{equation}
{\rm ln}A = \frac{1}{12}- \zeta^{(1)}(-1) = 0.248754477.
\end{equation}
The tensors we need are derived in \citet{1997JPhA...30.6485H}, except
that we have kept terms up to linear order in the expansion about
$K=0$ instead of dealing with the strictly wrenchless case.  Our
analysis therefore requires that $K \ll B_k^4$.
\begin{equation}
	\label{eqn:epsilon}
	\varepsilon^{i j} = \Delta^{i j} - \frac{\alpha}{2
          \pi}\left[\frac{2}{I} X_1\left(\frac{1}{\xi}\right)+
	\frac{12 K}{I^3}X_2\left(\frac{1}{\xi}\right)\right]B^i B^j
\end{equation}

\begin{equation}
	\label{eqn:mu}
	(\mu^{-1})^{ij} = \Delta^{i j} +  \frac{\alpha}{2
          \pi}\left[\frac{2}{I} X_1\left(\frac{1}{\xi}\right)+
	\frac{K}{12 I^3}X_2\left(\frac{1}{\xi}\right)\right]E^i E^j
\end{equation}
where

\begin{eqnarray}
	\Delta^{i j} &=& \delta^{i j}\Bigg [1 + \frac{\alpha}{2 \pi} \Bigg(-2 X_0\left(\frac{1}{\xi}\right) + \frac{1}{\xi}X_0^{(1)}\left(\frac{1}{\xi}\right) \nonumber \\ 
		& & + \frac{K}{4I^2}X_1\left(\frac{1}{\xi}\right) + \frac{K}{8 I^2\xi}X_1^{(1)}\left(\frac{1}{\xi}\right)\Bigg) \Bigg], 
\end{eqnarray}
the fine-structure constant is $\alpha=e^2/4 \pi$ in these
units where we have set $\hbar = c =1$,
and 
\begin{equation}
	\xi=\frac{1}{B_k}\sqrt{\frac{I}{2}}.
\end{equation}
Equations (\ref{eqn:epsilon}) and (\ref{eqn:mu}) define our model for
the QED vacuum in a strong electromagnetic field.  Because we will
focus on photon energies much lower than the rest-mass energy of the
electron, we have treated the vacuum as strictly non-linear.  It is
not dispersive.  The treatment of the dispersive properties of the
vacuum would require an effective action
treatment~\citep[e.g.][]{1995PhRvD..51.2513C,1995PhRvD..52.3163C}
rather than the local effective Lagrangian treatment used here.

\subsection{Plasma}
\label{sec:plasma}
To investigate travelling waves, we choose our coordinate system so
that the $\bhat{x}$-direction is aligned with the direction of
propagation.  Then, the spacetime dependence of the fields and sources
is given by a single parameter $S\equiv x-vt$ where $v$ is the
constant phase velocity in the $\bhat{x}$-direction of the travelling
wave.  At this point, we are choosing to work in a specific Lorentz
frame.

We model the plasma as a free electron plasma which enters the wave
equation through the source terms $\rho_p$ and $\Jp$.  For
electromagnetic fields obeying the travelling wave ansatz, the sources
must also obey the ansatz.  Then, $\rho_p$ and $\Jp$ are functions
only of $S$.  We therefore treat them as an additional field which is
integrated along with the electromagnetic components of the field.

In order to perform the numerical ODE integration for the currents, 
we wish to find expressions for $\dds{\rho_p}$
and $\dds{\Jp}$.  For travelling waves, the continuity 
equation is 
\begin{equation}
	\label{eqn:continuity}
	-v \deriv{\rho_p}{S} + \delta^{i x} \deriv{J_p^i}{S} = 0.
\end{equation}where we are using index notation to label our explicitly 
cartesian $\{x, y, z\}$ coordinate system.  Repeated indices are summed. However, 
whenever $x$ or $z$ appears as an index, it is 
fixed and does not run from 1 to 3.

We use equation (\ref{eqn:continuity}) to rewrite the source terms from equation (\ref{eqn:Maxwell1})
\begin{equation}
	\label{eqn:plasma1}
\partial^i \rho_p + \pderiv{{ J}^i_p}{t} = \frac{1}{v}\frac{d J_p^i}{d S}(\delta^{i x}-v^2).
\end{equation}Similarly, the source term from equation (\ref{eqn:Maxwell2}) is
\begin{equation}
	\label{eqn:plasma2}
(\curl{J_p})^i = \varepsilon^{ i x k} \left( \frac{d J_p}{dS} \right)^k.
\end{equation}

To find an expression for $\frac{d \Jp}{dS}$ we express the current as
an integral over the phase-space distribution of the electrons 
and linearize the plasma density.
\begin{equation}
	\label{eqn:Jpdef}
\Jp = \int{f({\bf p}_p) e {\bf v}_p d^3{\bf p}_p} \approx \bar{\gamma}ne \bar{\bf v}_p
\end{equation}where $\bar{\bf v}_p \equiv \frac{\int f({\bf p}_p) {\bf
    v}_p d^3 {\bf p}_p}{\bar{\gamma} n}$
and $n$ is the mean electron density in the plasma in the reference frame where $\bar{{\bf v}}_p=0$.
It is important to note the distinction between the mean plasma speed, $\bar{ v}_p$, and 
the propagation speed of the wave, $v$.
The Lorentz factor $\bar{\gamma}\equiv\frac{1}{\sqrt{1-\bar{v}_p^2}}$ accounts for a relativistic increase in the plasma density
 since $d^3 {\bf x} = \gamma^{-1} d^3 {\bf x}'$.

Next, we take a time derivative of the current and express this in terms of a
three-force acting on the plasma. 

 \begin{eqnarray}
 \label{eqn:dJdt2}
 \pderiv{{\bf J}_p}{t} &=& n e \pderiv{\bar{\gamma} \bar{{\bf v}}_p}{t} \nonumber \\
		& =&  \frac{n e}{m} {\bf F}
\end{eqnarray}

Noting that $\Jp$ is a function only of $S$, 
we insert the Lorentz force ${\bf F} = e(\efi + \bar{{\bf v}}_p \times \bfi)$
 and arrive at an expression that can be substituted into the source terms,
equations (\ref{eqn:plasma1}) and (\ref{eqn:plasma2}), to find $\Jp(S)$.

\begin{eqnarray}
\label{eqn:dJdS}
\frac{d {\bf J}_p}{dS} = -\frac{1}{v}\frac{e }{m} (e n\efi + \bar{\gamma}^{-1}{\bf J}_p \times \bfi)
\end{eqnarray}
The equations we have given above describe a cold, relativistic,
magnetohydrodynamic plasma.  Forces on the plasma arising due to
pressure gradients and gravity are neglected.  Moreover, in our
simulations, we neglect the forces on the plasma due to the magnetic
field.  This approximation is suitable if the plasma in question is a
pair plasma, or for wave frequencies much less than the cyclotron
frequency.  

As we will show in section \ref{sec:results} (see figure
 \ref{fig:freqvsspeedEz}), the field configurations 
generated in our simulation vary over timescales similar to the
inverse of the plasma frequency; the latter is 
about 9 orders of magnitude smaller than the 
cyclotron frequency.  This observation allows us to justify 
some aggressive assumptions regarding the plasma response.
As mentioned above, we may neglect forces on the plasma from 
the magnetic field, and quantum effects are expected to be 
small far away from 
the cyclotron resonance \citep{meszarosbook}.

We also neglect thermal effects since the influence of the 
electromagnetic fields will dominate over thermal motion.  
For strong background magnetic fields such that 
$\frac{eB}{m} \gg kT$, the electrons 
will be confined to the lowest Landau level, restricting 
thermal motion perpendicular to the background magnetic field.  
As we will elaborate on in section \ref{sec:results},
the greatest nonlinear effects occur for waves with an 
electric field component oriented along the background magnetic 
field.  Because we are interested in waves with amplitudes 
comparable to $B_k$, thermal motion is negligible 
relative to the dynamics induced by the wave.  
We are therefore justified in 
neglecting thermal effects in every direction
for the cases of greatest interest. 

If one combines Eq.~(\ref{eqn:dJdt2}) with
Eq.~(\ref{eqn:Maxwell1}), one sees that any nonlinearity in this
treatment must originate with the dielectric and permeability tensors
--- any non-linearity that may originate from the plasma itself has been
neglected \citep[c.f.][]{1979JETP...49...75K,cattaert:012319}. The
plasma is modelled as strictly dispersive. 
In section~\ref{sec:WFL},
we confirm that this method of describing the plasma is consistent
with standard accounts in the weak-field, small-wave limit.  

\subsection{Travelling Wave ODEs}
 
The wave equations for travelling waves, are found by combining
Maxwell's equations (\ref{eqn:Maxwell1}) and (\ref{eqn:Maxwell2}) with
the continuity equation, (\ref{eqn:continuity}).  We also make the
plane-wave approximation, and assume that the fields and sources are
described by the parameter

\begin{equation}
	\label{eqn:S}
	S=x-vt.
\end{equation}
The equations governing travelling wave propagation are

\begin{equation}
\label{eqn:ODE1}
\ddS{\psi^i(S)}=\frac{1}{v}\dds{J^i(S)}(\delta^{i x} - v^2)
\end{equation}

\begin{equation}
\label{eqn:ODE2}
\ddS{\chi^i(S)}=-\varepsilon^{i x j} \dds{J^j(S)}
\end{equation}
  The 
auxiliary vectors $\psi^i$ and $\chi^i$ are related to the electric and magnetic fields.
\begin{eqnarray}
\label{eqn:u}
\psi^i(S)&=&(1-v^2)\varepsilon^{i j} E^j + \delta^{ix}(\varepsilon^{x j} E^j-E^x)- \nonumber \\
& & ~~(\varepsilon^{i j}E^j-E^i) + \varepsilon^{i x
  k}v(B^k-(\mu^{-1})^{k j}B^j) \\
\label{eqn:v}
\chi^i(S)&=&((\mu^{-1})^{i j} B^j-v^2 B^i) + \delta^{ix}(B^x-(\mu^{-1})^{x j}B^j) - \nonumber \\
& & ~~\varepsilon^{i x k}v(\varepsilon^{k j}E^j-E^k)
\end{eqnarray}
Equations (\ref{eqn:ODE1}) through (\ref{eqn:v}) define a set of
coupled ordinary differential equations that can be integrated to
solve for the travelling electric and magnetic fields.  Solving these
equations requires that we have at hand the vacuum dielectric and
inverse magnetic permeability tensors as well as an expression for
$\frac{d \Jp}{d S}$.  These were discussed in sections
\ref{sec:tensors} and \ref{sec:plasma} respectively.

\subsection{Weak field, Small Wave limit}
\label{sec:WFL}
In this section, we would like to demonstrate that our equations
reduce to standard expressions in the case of small background fields
and small electromagnetic waves.  To make this comparison, it is also
prudent to assume waves have a spacetime dependence like $e^{i(\omega
  t - {\bf k}\cdot{\bf x})}$ instead of $x-v t$.  By making this
change, we can compare our other assumptions with those made in
standard textbook accounts directly.

Under the assumption that the fields and currents have the standard
plane-wave spacetime dependence, we can make the replacements
$\pderiv{}{t} \rightarrow -i\omega$ and $\nabla \rightarrow i {\bf
  k}$.  We can then write a second expression for $\pderiv{{\bf
    J}}{t}$.

\begin{equation}
\label{eqn:dJdt1}
\pderiv{\Jp}{t} = -i \omega \Jp
\end{equation}

Setting equations (\ref{eqn:dJdt1}) and (\ref{eqn:dJdt2}) equal, we get

\begin{equation}
\label{eqn:ainJ}
\frac{e}{m}(\varepsilon^{i j k} J^j_p B^k) + i \omega J^i_p = \frac{e^2}{m} n E^i
\end{equation}where we have used the nonrelativistic approximation, which is appropriate
for small waves.  We will now specialize to a background magnetic field pointing in 
the $\bhat{z}$-direction.  We can then write this background field in terms of the 
cyclotron frequency $\omega_c = \frac{eB}{m}$.  The density $n$ determines the 
plasma frequency $\omega_p^2 =n \frac{e^2}{m}$.  Then, we can write 
equation (\ref{eqn:ainJ}) as
\begin{equation}
\label{eqn:ainJ2}
\left[ -\omega_c \varepsilon^{i j z} -  i \omega \delta^{j i}\right] J_p^j = \omega_p^2 E^i
\end{equation}

We can solve this for $\Jp$ by writing a matrix equation

\begin{equation}
\Jp =  \left( \begin{array}{ccc}
-i \omega &  -\omega_c & 0 \\
\omega_c & -i\omega & 0 \\
0 & 0 & -i\omega \end{array} \right)^{-1} 
\omega_p^2 {\bf E}
\end{equation}inverting the matrix gives
\begin{equation}
\label{eqn:Jmatrix}
\Jp =  \left( \begin{array}{ccc}
\frac{i \omega}{\omega^2-\omega_c^2} & \frac{-\omega_c}{\omega^2-\omega_c^2} & 0 \\
\frac{\omega_c}{\omega^2-\omega_c^2} & \frac{i \omega}{\omega^2-\omega_c^2} & 0 \\
0 & 0 & i/\omega \end{array} \right) 
\omega_p^2 {\bf E}.
\end{equation}

Finally, we would like to use equation (\ref{eqn:Jmatrix}) to write the right hand side of 
equation (\ref{eqn:Maxwell1}) in the small wave limit in a manner we can interpret as 
a dielectric tensor.

Ignoring (for now) the contributions from the vacuum, 
and assuming an approximately homogeneous plasma density, 
equation (\ref{eqn:Maxwell1}) simplifies to
\begin{eqnarray}
\label{eqn:RHSMax1}
\laplace {\bf E} - \ddt{{\bf E}} &=& \pderiv{\Jp}{t} \nonumber \\ 
	&=& - i\omega\Jp .
\end{eqnarray}
If our macroscopic field is to obey $\laplace {\bf D} - \ddt{{\bf D}}=0$,
we can insert equation (\ref{eqn:Jmatrix}) into (\ref{eqn:RHSMax1}) to obtain the 
following expression for the dielectric tensor due to plasma effects:

\begin{equation}
\varepsilon^{(p)}_{i j} = \left( \begin{array}{ccc}
1-\frac{\omega_p^2}{\omega^2-\omega_c^2} & -i\frac{\omega_c}{\omega}\frac{ \omega_p^2}{\omega^2-\omega_c^2} & 0 \\
i\frac{\omega_c}{\omega}\frac{ \omega_p^2}{\omega^2-\omega_c^2} & 1-\frac{\omega_p^2}{\omega^2-\omega_c^2} & 0 \\
0 & 0 & 1-\left(\frac{\omega_p^2}{\omega^2}\right) \end{array} \right)
\label{eq:2}
\end{equation}This expression is in agreement with the cold plasma dielectric tensor given in \citet{meszarosbook}.
As noted above, our analysis neglects the off-diagonal (Hall) terms, as is appropriate for pair plasmas 
or waves with frequencies much less than the cyclotron frequency.

Because the vacuum effects are added explicitly in the form of dielectric and magnetic permeability tensors,
we only need to confirm that the weak field limits of our expressions agree with the standard results.
This confirmation is done explicitly in \citet{1997JPhA...30.6485H}.

In the weak field limit, the tensors given by equations (\ref{eqn:epsilon}) 
and (\ref{eqn:mu}) are
\begin{eqnarray}
\varepsilon_{ij}^{(v)} &=& \delta_{ij} + \frac{1}{45 \pi} \frac{\alpha}{B_k^2}
\left [ 2(E^2-B^2) \delta_{ij} + 7 B_i B_j \right ], \\
\mu^{-1 (v)}_{ij} &=& \delta_{ij} + \frac{1}{45 \pi} \frac{\alpha}{B_k^2}
\left [ 2(E^2-B^2) \delta_{ij} - 7 E_i E_j \right ] 
\end{eqnarray}
  In the case of a weak background magnetic field pointing in the $\bhat{z}$-direction,
these become
\begin{eqnarray}
\varepsilon_{ij}^{(v)} &=& \left( \begin{array}{ccc}
1-2\delta & 0 & 0 \\
0 & 1-2\delta & 0 \\
0 & 0 & 1+5\delta  \end{array} \right)
\label{eq:1}
\\
\mu_{ij}^{-1 (v)} &=& \left( \begin{array}{ccc}
1-2\delta & 0 & 0 \\
0 & 1-2\delta & 0 \\
0 & 0 & 1-6\delta \end{array} \right)
\end{eqnarray}with 

\begin{equation}
\label{eqn:delta}
\delta = \frac{\alpha}{45 \pi} \left( \frac{B}{B_k} \right) ^2
\end{equation}Again, this result agrees with the vacuum tensors given in \citet{meszarosbook}.

In this limit, we may simply add together the contributions to the dielectric tensor 
from the plasma and the vacuum according to

\begin{equation}
\varepsilon_{i j} = \delta_{i j} + (\varepsilon_{i j}^{(p)}-\delta_{i j}) + (\varepsilon_{i j}^{(v)}-\delta_{i j})
\end{equation} with $\mu_{i j}^{-1}$ given entirely by the vacuum contribution.

We have thus recovered the standard result for a medium consisting of a plasma and the QED
vacuum in the weak field, small wave limit.

\section{Solution Procedure}
In total, there are 15 coupled non-linear ODEs which must be
integrated to produce a solution.  Equations (\ref{eqn:ODE1}) and
(\ref{eqn:ODE2}) define the electric and magnetic fields as functions
of $S$.  In addition, we must simultaneously integrate equation
(\ref{eqn:dJdS}) which gives the plasma current as a function of $S$.
Each of these equations has three spatial components.
Initial conditions are given for each of the six field components, and
the six derivatives of the field with respect to $S$.  The equations
describing the electromagnetic fields do not depend on the initial
values of the current, but in principle these are the three remaining
initial conditions.

\begin{figure}
	\centering
		\includegraphics[width=8.5cm]{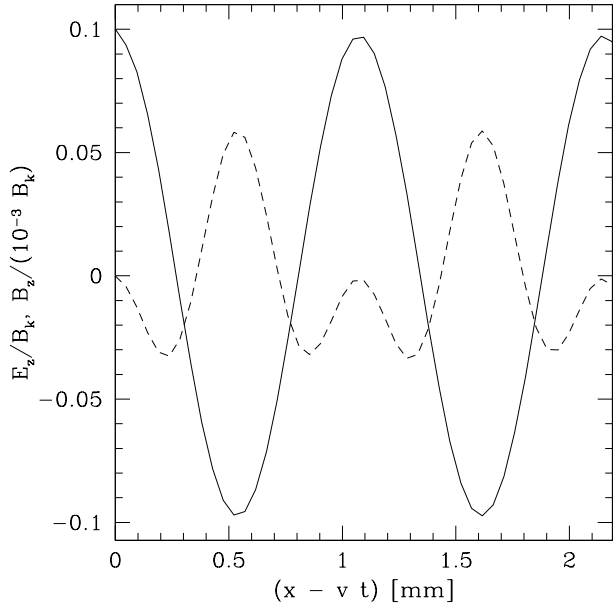}
                \caption{A comparison between the $\bhat{z}$
                  components of the electric (solid) and magnetic
                  fields (dashed) showing the nonorthogonal
                  stabilizing wave. On larger scales, the $B_z$
                  component is seen to have a periodic envelope
                  structure as in figure \ref{fig:envelope}}.
	\label{fig:EzBzcompare}
\end{figure}

\begin{figure}
	\centering
		\includegraphics[width=8.5cm]{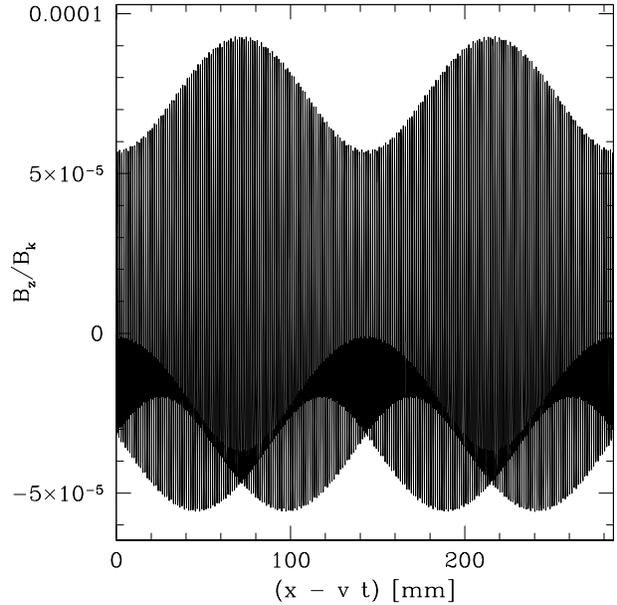}
                \caption{The $B_z$ wave component as shown
                  in figure \ref{fig:EzBzcompare} has an envelope
                  structure when viewed at larger scales.  In this
                  case, the $E_z$ wave component has a nearly constant
                  amplitude of $0.1 B_k$.}.
	\label{fig:envelope}
\end{figure}
The ODEs are solved using a variable stepsize Runge-Kutta method.  In
order to translate between the $\efi$-and-$\bfi$ picture and the ${\bf
  \psi}$-and-${\bf \chi}$ picture, equations (\ref{eqn:u}) and
(\ref{eqn:v}) must be solved numerically at each time step, including
the first step when the initial conditions are given.

Tables of numerical values of the functions defined by equations
(\ref{eqn:x0anal}) to (\ref{eqn:x2anal}), as well as their
derivatives, were computed in advance and these were used to
interpolate the values needed in the simulation using a standard cubic
spline interpolation algorithm.  In producing these tables,
expressions for the weak and strong field limits were used in the
appropriate regimes as this reduced the numerical errors.

Aside from the initial conditions for the fields and derivatives,
there is one parameter in the model which must be selected.  The
density of the plasma, given by $n$ in equation(\ref{eqn:Jpdef}) is
chosen to be
\begin{equation}
n = 10^{13}\mathrm{cm}^{-3}.
\end{equation}
This value corresponds to the Goldreich-Julian density for a star of
period $P\sim 1$~s, $\dot{P}\sim 10^{-10}$.  The uniform background
field is taken to equal the quantum critical field strength 
\begin{equation}
	B_k = \frac{m^2}{e} = 4.413 \times 10^{13} \rmmat{G}.
\end{equation}

\section{Results}
\label{sec:results}

This study focuses on the case of waves propagating transverse to a
large background magnetic field.  We have already chosen the direction
of propagation to be the $\bhat{x}$-direction through our definition
of $S$ in Eq.~(\ref{eqn:S}).  We now choose the background magnetic
field to point in the $\bhat{z}$-direction.  In this situation, the
largest non-linear effects occur when there is a large amplitude wave
in the $\bhat{z}$ component of the electric field.  This is quite
natural.  The values of both the dielectric tensor of the plasma
(Eq.~\ref{eq:2}) and the weak-field limit of the vacuum dielectric
tensor (Eq.\ref{eq:1}) differ most from unity for this component; for
strong fields the index of refraction (as well as its derivative with
respect to the field strength) is largest for vacuum propagation in
this mode \citep{1997JPhA...30.6485H}.  Furthermore, the dominant
three-photon interaction couples photons with the electric field
pointed along the global magnetic field direction with photons whose
magnetic field points along this direction.  The three-point
interaction for photons whose electric field is perpendicular to the
magnetic field with parallel photons vanishes by the $CP-$invariance
of QED~\citep{Adle71}.  Therefore, we focus on initial
($S=0$) conditions in which the dominant component of the electric
field points along in the $\bhat{z}$-direction and that of the
magnetic field along the $\bhat{y}$-direction.

In the classical vacuum, these initial conditions correspond to
transverse, linearly polarized sine-wave solutions that travel at the
speed of light.  However, when the wave amplitudes are large, and there
is a strongly magnetized plasma, we find that there is a deviation from
normal transverse electromagnetic waves.  In particular, in order to
remain stable, a wave with large $E_z$ and $B_y$ field components must
also excite waves in the $E_y$ and $B_z$ fields.  These stabilizing
wave components exhibit strong non-linear characteristics (see figures
\ref{fig:EzBzcompare} and \ref{fig:envelope}).  The symmetries of the
wave equation require a close correspondence between the $E_y$ and
$B_z$ waveforms as well as between the $B_z$ and $B_y$ waveforms.  For
simplicity, only one of each is plotted in the figures.  The field
strengths are given in units of the quantum critical field strength,
$B_k$.

As is apparent from Fig.~\ref{fig:EzBzcompare} the dominant electric
field along the direction of the external magnetic field is
essentially sinusoidal.  Subsequent figures will show that there is a
small harmonic component.  The waveform for the dominant magnetic
field component is similar.  On the other hand, the magnetic field
along the direction of the electric field (the non-orthogonal
component) is smaller by nearly four orders of magnitude and obviously
exhibits higher harmonics.  In particular if one expands the scale of
interest (Fig.~\ref{fig:envelope}), the magnetic field exhibits
beating between two nearby frequencies with similar power.

In order to examine the harmonic content of the waveforms, we perform
fast Fourier transforms (FFTs) on the signals produced in the
simulations.  We present the results in terms of power spectra
normalized by the square amplitudes of the electric field of the
waves.  In these plots, the horizontal axis is normalized by the
frequency with the greatest power in the electric field, so that
harmonics can be easily identified.

\begin{figure}
	\centering
		\includegraphics[width=8.5cm]{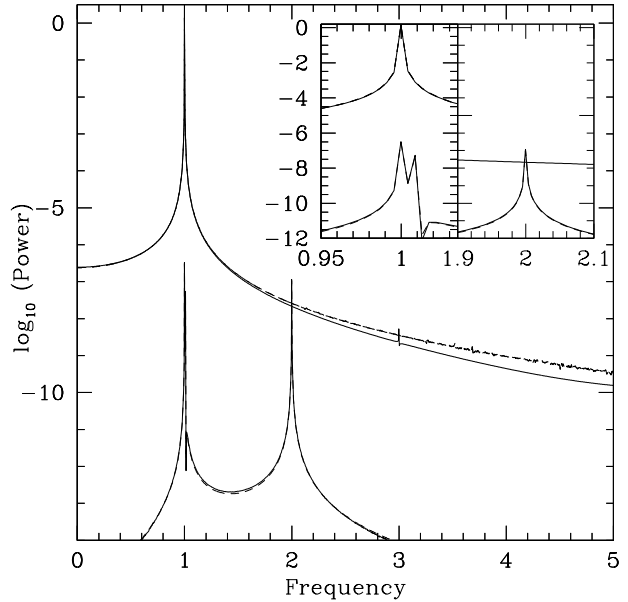}
                \caption{The solid curve depicts power spectra of the
                  electric (upper) and magnetic fields (lower) along the
                  $\bhat{z}$-direction for the solution depicted in
                  Figs.~\ref{fig:EzBzcompare} and~\ref{fig:envelope}.
                  The inset focusses on the fundamental and the first
                  harmonic.  The dashed curve follows the power
                  spectrum of a single sinusoid for the electric field
                and three sinusoids for the magnetic field. Near the
                peaks the dashed curve is essentially
                indistinguishable from the solid one.}
	\label{fig:PScompare}
\end{figure}

Fig.~\ref{fig:PScompare} depicts the power spectra of the electric and
magnetic fields along the $\hat{z}$-direction for the wave depicted in
Figs.~\ref{fig:EzBzcompare} and~\ref{fig:envelope}.  The conclusions
gathered from an examination of the waveforms are born out by the power
spectra.  In particular the electric field is a pure sinusoidal
variation to about one part in ten thousand -- the power spectrum of a
pure sinusoid is given by the dashed curved.  The duration of the
simulation is not an integral multiple of the period of the sinusoid,
resulting in a broad power spectrum even for a pure sinusoid.
The power spectrum of the magnetic field follows the expectations
gleaned from the waveforms.  In particular the fundamental and the
first harmonic are dominant, with the first harmonic having about
one-third the power of the fundamental.  If one focusses on the
fundamental, one sees that two frequencies are involved.  The envelope
structure is produced by a beating between the fundamental and a
slightly lower frequency with a similar amount of power as the first
harmonic.   Over the course of the simulation the envelope exhibits
two apparent oscillations; the lower frequency differs by two
frequency bins, so it is resolved separately from the fundamental as
shown in the inset.

\begin{figure}
	\centering
		\includegraphics[width=8.5cm]{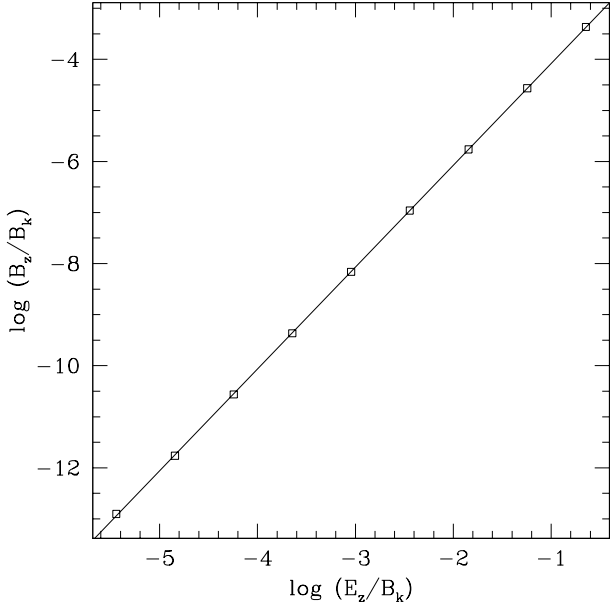}
                \caption{Amplitude of the $B_z$ component plotted against the amplitude of the 
		$E_z$ component for a $B_0 = B_k$ background field.
                The line is the best-fitting power-law relation. The slope is consistent with 
		a scaling exponent equal to two.
              }
	\label{fig:ampvsamp}
\end{figure}

As the amplitude of the electric field increases the non-linear and
non-orthogonal features of the travelling wave increase.
Fig.~\ref{fig:ampvsamp} shows that the strength of the non-orthogonal
magnetic field increases as the square of the electric field, a
hallmark of the non-linear interaction between the fields.  For the
strongest waves studied with $E_z \approx 0.2 B_k$ (the rightmost
point in the figure), the magnetic field, $B_z$, is about $10^{-4} B_k$ nearly
one-percent of the electric field.  The amplitude of the
non-orthogonal magnetic field is given by
\begin{equation}
B_z = 0.008 B_k \left ( \frac{E_z}{B_k} \right )^2
\label{eq:3}
\end{equation}
for $B_0=B_k$.  The coefficient is coincidentally very close to
three-quarters of the value of the fine-structure constant.  It
increases with the strength of the background field.

\begin{figure}
	\centering
		\includegraphics[width=8.5cm]{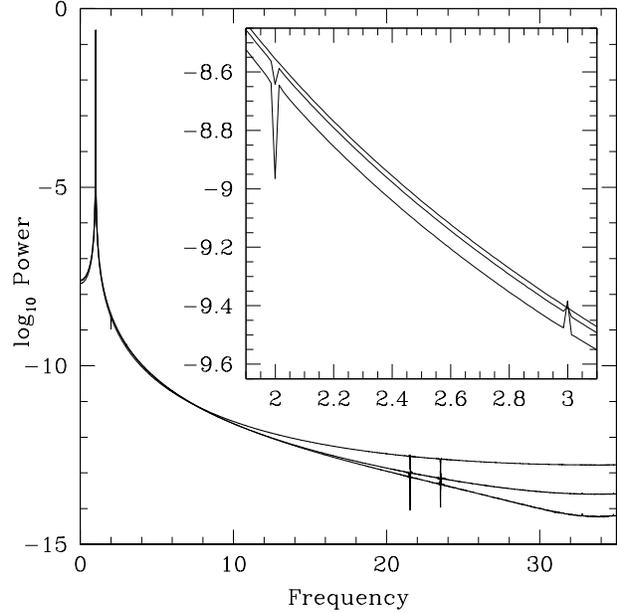}
                \caption{This power spectrum demonstrates the
                  development of nonlinear effects in the
                  $\bhat{z}$-component of the electric field as the
                  amplitude of the wave is increased in a $B_0=B_k$
                  background field. From top to bottom the curves
                  follow the solutions whose amplitude of $E_z$ equals
                  $0.01, 0.08$ and $0.16 B_k$.  The inset focusses on the first
                  and second harmonics.}
	\label{fig:PSvsampE}
\end{figure}

For the strongest waves even the non-orthogonal magnetic field is
strong, so it can generate non-linearities in the electric field.
Although the strongest effect is around the fundamental, it is
completely swamped by the fundamental of the electric field.  On the
other hand, the magnetic field drives a first and second harmonic in
the electric field as seen in Fig.~\ref{fig:PSvsampE}.  The strength
of these harmonics is approximately given by the formula in
Eq.~(\ref{eq:3}) or equivalently Fig.~\ref{fig:ampvsamp} if one
substitutes the value of $B_z$ for $E_z$ and uses result for $E_z$.
This is essentially a six-order correction from the effective
Lagrangian.  Because we have used the complete Lagrangian rather than
a term-by-term expansion, all of the corrections up to sixth order
(and further) are automatically included in the calculation.

\begin{figure}
  	\centering
		\includegraphics[width=8.5cm]{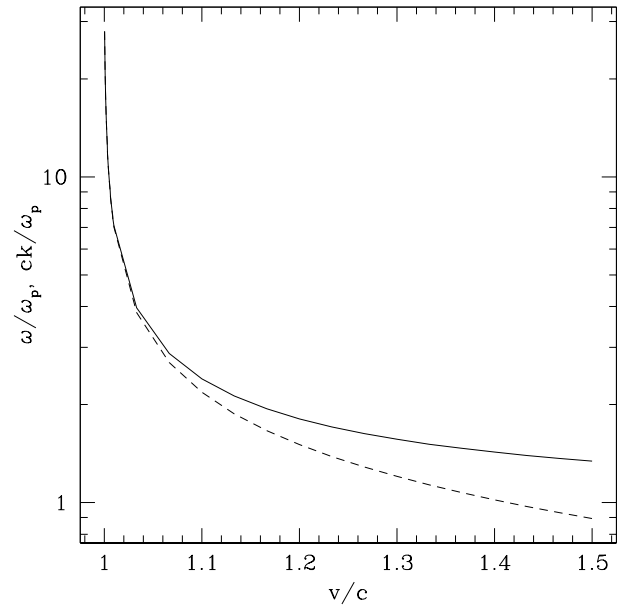}
                \caption{The frequency (solid curve) and wavenumber
                  (dashed curve) of the travelling wave as a
                  function of its phase velocity.  }
	\label{fig:freqvsspeedEz}
\end{figure}

Figure \ref{fig:freqvsspeedEz} demonstrates how the frequency of the
solutions varies with the speed of propagation.  Because there is no
mode information stored directly in our numerical solutions, we take
the frequency to be rate that local minima in the electric field pass a
fixed observer.  In general, we find that the frequency of traveling
waves increases as the phase velocity approaches the speed of light,
very closely following the formula
\begin{equation}
\left ( \frac{\omega}{\omega_p} \right )^2 =  \frac{v^2}{v^2-1}.
\end{equation}
This formula also results from an analysis of the dielectric tensor,
Eq.~(\ref{eq:2}).  The vacuum makes a small contribution to the wave
velocity in this regime.
%

\section{Conclusion}

We have discussed techniques for computing electromagnetic waves in a
strongly magnetized plasma which nonperturbatively account for the
field interactions arising from QED vacuum effects.  We
applied these methods to the case of travelling waves, which have a
spacetime dependence given by the parameter $S=x-vt$.  Travelling
waves can be described without any decomposition into Fourier modes
and this is ideal for exploring the nonperturbative aspects of waves.

The main result from this analysis is the observation that
electromagnetic waves in a strongly magnetized plasma can
self-stabilize by exciting additional nonorthogonal wave components.
In the cases studied above, a large amplitude excitation of the
electromagnetic field, for example, from the coupling between Alfv\'en
waves to starquakes, can induce nonlinear waves which are stabilized
against the formation of shocks.  The result is a periodic wave train
with distinctly nonlinear characteristics.  Such structures may play a
role in forming pulsar microstructures.

This result demonstrates that shock formation is not a necessary outcome 
for waves in a critically magnetized plasma.  It is possible that 
nonlinear features of a wave can stabilize it against shock formation. 
The shock-wave solutions generally decrease in magnitude, dissipating 
energy along the way unless they travel into a low-field region, so the 
self-stabilizing nonlinear waves are the 
only ones that keep their shape and energy content intact as they propagate.
It is not yet clear what set of conditions will determine 
if a particular wave will self-stabilize or collapse to form a shock or how 
plasma inhomogenaities will affect the propagation of a travelling 
wave train.   These are issues which can be clarified in future work.

\bibliographystyle{mn2e}
\bibliography{references}
\label{lastpage}
\end{document}